\documentstyle[prd,aps,multicol,epsfig]{revtex}

\renewcommand{\narrowtext}{\begin{multicols}{2}\global\columnwidth20.5pc}
\renewcommand{\widetext}{\end{multicols}\global\columnwidth42.5pc}
\begin{document}
\draft
\title{Fake symmetry transitions in lattice Dirac spectra}
\author{M. Schnabel$^1$ and T. Wettig$^2$}
\address{$^1$Institut f\"ur Theoretische Physik,
  Universit\"at Regensburg, D-93040 Regensburg, Germany\\
  $^2$Center for Theoretical Physics, Yale University, New Haven,
  CT 06520-8120}  
\date{printed \today}
\maketitle
\begin{abstract}
  In a recent lattice investigation of Ginsparg-Wilson-type Dirac
  operators in the Schwinger model, it was found that the symmetry
  class of the random matrix theory describing the small Dirac
  eigenvalues appeared to change from the unitary to the symplectic
  case as a function of lattice size and coupling constant.  We
  present a natural explanation for this observation in the framework
  of a random matrix model, showing that the apparent change is caused
  by the onset of chiral symmetry restoration in a finite volume.  A
  transition from unitary to symplectic symmetry does not occur.
\end{abstract}

\narrowtext

\section{Introduction}
\label{sec1}

In the past two years, Dirac operators ${\mathcal D}$ satisfying the
Ginsparg-Wilson (GW) condition \cite{GW82},
\begin{equation} 
  {\mathcal D}\gamma_5+\gamma_5{\mathcal D}= 
  2{\mathcal D}\gamma_5R{\mathcal D}\:,
  \label{GWC}
\end{equation}
have attracted a great deal of attention in the lattice community
because of their vastly improved chiral and topological properties.
In a recent lattice study of the Schwinger model, Farchioni et al.
\cite{FHLW98a} investigated two different versions of ${\mathcal D}$,
a fixed point Dirac operator ${\mathcal D}_{\rm fp}$ \cite{Has98a} and
Neuberger's Dirac operator ${\mathcal D}_{\rm Ne}$ \cite{Neu98a}.  For
the purpose of the present study, these two operators have essentially
identical properties, and we concentrate on ${\mathcal D}_{\rm Ne}$ in
the following.  The subtracted lattice condensate is defined by
\cite{Neu98a,Has98a}
\begin{equation}
  \langle\bar{\psi}\psi\rangle_{\rm sub}=-\frac{1}{V}\left\langle
    {\rm tr}\:\tilde{\mathcal D}^{-1}\right\rangle_{\rm gauge}\:,
\label{condensate} 
\end{equation}
where $\tilde{\mathcal D}={\mathcal D}(1-R{\mathcal D})^{-1}$ and $V$
is the physical volume.  The operator $\tilde{\mathcal D}$ shares some
important features with the continuum Dirac operator in Euclidean
space.  It is anti-Hermitian and satisfies $\{\tilde{\mathcal
  D},\gamma_5\}=0$.  Thus, its nonzero eigenvalues occur in pairs $\pm
i{\lambda_n}$ with $\lambda_n$ real.  The spectral density of
$i\tilde{\mathcal D}$ is defined by
$\rho(\lambda)=\langle\sum_n(\lambda-\lambda_n)\rangle/V$.  The
Banks-Casher relation \cite{Banks80} then reads
\begin{equation}
\label{BC}
  \Sigma\equiv-\lim_{V\to\infty}\langle\bar{\psi}\psi\rangle_{\rm sub}
  =\lim_{\lambda\to0}\lim_{V\to\infty}\pi\rho(\lambda)\:,
\end{equation} 
where it is important that the thermodynamic limit is taken first.  

Our main concern in the present work are the somewhat puzzling
observations made in Ref.~\cite{FHLW98a} where the spectral properties
of $\tilde{\mathcal D}$ were compared to predictions of chiral random
matrix theory (RMT) \cite{Shur93}.  RMT is a simple model which yields
exact analytical results for the spectral correlations of the Dirac
operator on the scale of the mean level spacing.  Because of the
chiral structure of the problem, one has to distinguish two regions in
the spectrum, (i) the bulk and (ii) the eigenvalues in the vicinity of
zero (the latter is called the microscopic region or the hard edge of
the spectrum).  The analytical predictions of RMT are different in
these two regions.  Furthermore, there are three different symmetry
classes, the chiral orthogonal (chOE), unitary (chUE), and symplectic
(chSE) ensembles \cite{Verb94}.  

Let us briefly summarize those findings of Ref.~\cite{FHLW98a} which
are of relevance to the present work.  The Schwinger model (QED$_2$)
is in the symmetry class of the chUE.  For large volumes $V\propto
L^2/\beta$ ($L$ is the number of lattice sites in each dimension and
$\beta=1/(ea)^2$ is the dimensionless coupling), it was found that the
bulk as well as the hard edge of the spectrum are nicely described by
the chUE predictions.  As $V$ decreases, the bulk properties are still
given by the chUE whereas at the hard edge, the data suggest a
transition to chSE behavior.  In Ref.~\cite{FHLW98a}, the sectors of
topological charge $\nu=0$ and $\nu=1$ were investigated, and the
apparent transition was seen in both sectors.  These observations are
puzzling, since it is not clear where the symplectic symmetry should
come from.  Also, the spectrum is not doubly degenerate as expected
for the chSE.  The authors of Ref.~\cite{FHLW98a} already suggested
that their observation might be an artifact of the small physical
volume and that the agreement with the chSE may be accidental.

In the framework of RMT, a transition between the two symmetry classes
can be described by the simple model
\begin{equation} 
  (1-\alpha)\,{\mathcal D}_{\rm chUE}+\alpha\,{\mathcal D}_{\rm chSE}
  \label{first}
\end{equation}
with Dirac operators of the appropriate symmetries and
$0\le\alpha\le1$ interpolating between the two ensembles.  However,
in such a model the transition from chUE to chSE is expected to occur
simultaneously in the bulk and in the microscopic domain.  We have
confirmed this expectation numerically.  Thus, the simple ansatz
(\ref{first}) cannot explain the findings of Ref.~\cite{FHLW98a}.
This is as it should be, since we do not expect a symplectic symmetry
to be present in the first place.

For RMT to be applicable at the hard edge, it is necessary that chiral
symmetry is spontaneously broken, i.e., we require $\rho(0)>0$.
(Since on the lattice one is always working at finite volume, we mean
here the value obtained by extrapolating $\rho(\lambda)$ many level
spacings away from zero to $\lambda=0$.)  As we shall see, the
apparent transition from chUE to chSE symmetry in the microscopic
domain is caused by the fact that $\rho(0)$ vanishes as the physical
volume decreases.  This will be shown in more detail below, using a
random matrix model which we construct in the following section.  The
spectral properties we consider are introduced in Sec.~\ref{sec3}, our
results are presented and discussed in Sec.~\ref{sec4}, and
conclusions are drawn in Sec.~\ref{sec5}.

\section{The random matrix model}
\label{sec2}

Our approach closely parallels the construction of the Neuberger
operator \cite{Neu98a}.  The Wilson Dirac operator for the $d=2$
dimensional Schwinger model defined on an $L\times L$ lattice reads
\begin{eqnarray}
  {\mathcal M}_{x,y}=\delta_{x,y}-\kappa\sum_{\mu=1}^2
  \big[&&(1-\sigma_\mu)U_\mu(x)\Delta_{x,y-\hat\mu}+\nonumber\\
  &&(1+\sigma_\mu)U^\dagger_\mu(x-\hat\mu)\Delta_{x,y+\hat\mu}\big]  
  \label{Wilson}
\end{eqnarray}
with hopping parameter $\kappa=1/(2m+4)$ and Pauli matrices
$\sigma_i$.  As in Ref.~\cite{FHLW98a}, we will restrict ourselves to
the case $m=-1$, $\kappa=1/2$.  The boundary conditions for the
fermions are periodic in space and anti-periodic in Euclidean time.
This is taken into account by the function $\Delta_{x,y\pm\hat\mu}$
which is the usual Kronecker delta $\delta_{x,y\pm\hat\mu}$ except for
the links from $x_2=L$ to $x_2=1$ for which there is an extra factor
of $-1$.  The U(1) gauge fields are represented by $2L^2$ phases
$U_\mu(x)$ which obey periodic boundary conditions and fluctuate in
the update process.  In our model, they are simply replaced by
independent uncorrelated random phases $\exp(i\varphi)$ with $\varphi$
drawn at random from the interval $[-\delta,\delta]$.  For a single
flavor with two spinor indices, $\mathcal M$ thus becomes a (sparse)
random matrix of dimension $2L^2$.

{}From the matrix $\mathcal W$ obtained by replacing the gauge fields
by random phases we construct the Neuberger operator \cite{Neu98a},
\begin{equation}
  {\mathcal W}_{\rm Ne}=\openone+
  \gamma_5\,\varepsilon(\gamma_5{\mathcal W})\:,
\end{equation}
where $\varepsilon$ is the sign function $\varepsilon(A)\equiv
A/\sqrt{AA^\dagger}$.  ${\mathcal D}_{\rm Ne}$ (and hence also
${\mathcal W}_{\rm Ne}$) satisfies the GW condition (\ref{GWC}) with
$R=1/2$. Its spectrum is located on the unit circle in the complex
plane with center at $z=1$.  Furthermore, for certain background
fields it possesses exact zero modes whose number is related to the
topological charge of the background field.  In analogy with the
definition of $\tilde{\mathcal D}$ we finally construct the random
matrix
\begin{equation}
  \widetilde{\mathcal W}={\mathcal W}_{\rm Ne}
  (1-{\mathcal W}_{\rm Ne}/2)^{-1}\:.
\end{equation}
This projects the spectrum of ${\mathcal W}_{\rm Ne}$ onto the
imaginary axis.  In the following, we will investigate the spectrum of
the matrix $\widetilde{\mathcal W}$.  

There are two parameters in the random matrix model, the linear
lattice size $L$ and the width $2\delta$ of the distribution
$P(\varphi)=\theta(\delta+\varphi)\theta(\delta-\varphi)/2\delta$.
The parameter $\delta$ is restricted to the interval
$0\le\delta\le\pi$.  It models the coupling $\beta$ used in the
lattice gauge simulation.  Small values of $\delta$ correspond to
phases $\exp(i\varphi)$ which fluctuate only weakly around unity.
Thus, small values of $\delta$ correspond to large values of $\beta$
and vice versa.

An analytical treatment of the problem is very difficult.  Therefore,
we performed a numerical study for several combinations of the
parameters $L$ and $\delta$.  As we shall see, the random matrix model
is able to reproduce and to explain the findings of
Ref.~\cite{FHLW98a}.

\section{Spectral properties}
\label{sec3}

Since our main purpose in this work is to understand the puzzling
observations of Ref.~\cite{FHLW98a}, we have investigated the same
quantities considered therein.  These are the spectral density
$\rho(\lambda)$, the distribution of the smallest (positive)
eigenvalue $P(\lambda_{\rm min})$, and the nearest neighbor spacing
distribution $P(s)$.  The quantity $s$ is the spacing between adjacent
eigenvalues in the bulk after the spectrum has been unfolded
\cite{Meht91}.  $P(\lambda_{\rm min})$ and $P(s)$ measure the spectral
properties at the hard edge and in the bulk of the spectrum,
respectively.  (We will not discuss the microscopic spectral density
which was also considered in Ref.~\cite{FHLW98a} since for the purpose
of the present study, $P(\lambda_{\rm min})$ is sufficient to reveal
the properties of the spectrum near $\lambda=0$.)  The global density
$\rho(\lambda)$ of the random matrix model is not a universal
quantity, i.e., it depends on the details of the model and is not
expected to agree with real lattice data.  However, $P(\lambda_{\rm
  min})$ and $P(s)$ are universal for a specific symmetry class, i.e.,
these functions are insensitive to the details of the dynamics, and
the results obtained in RMT should agree with those of lattice
simulations (unless the conditions are such that RMT is not
applicable, see below).

The nearest neighbor spacing distribution $P(s)$ in the bulk of the
spectrum is expected to agree with the RMT result irrespective of
whether or not chiral symmetry is broken \cite{Pull98}.  In other
words, it is insensitive to the value of $\rho(0)$.  $P(s)$ can be
unambiguously constructed from the lattice data without any free
parameter by unfolding.  The RMT results for $P(s)$ are well
approximated by the Wigner surmise \cite{Meht91}
\begin{equation}
  \label{Ps}
  P(s)=\left\{
      \begin{array}{ll}
      (32/\pi^2)\,s^2\exp\left(-4s^2/\pi\right)  & {\rm chUE}\\
      (2^{18}/3^6\pi^3)\,s^4\exp\left(-64s^2/9\pi\right) & {\rm chSE}\:.
    \end{array}  
  \right. 
\end{equation}
The nearest neighbor spacing $s$ is expressed in units of the mean
level spacing in the bulk.  Note that $P(s)$ does not depend on
$\nu$ or on the number of massless flavors $N_f$.

In contrast to $P(s)$, $P(\lambda_{\rm min})$ is only given by the RMT
result if chiral symmetry is broken, i.e., if $\rho(0)\ne0$.  Also in
contrast to $P(s)$, it depends on the number of zero modes $\nu$ and
on $N_f$.  In the following, we restrict ourselves to the quenched
case, $N_f=0$.  The energy scale for a comparison between the RMT
result and numerical data is set by the mean level spacing at the hard
edge, $1/V\rho(0)=\pi/V\Sigma$, which can be determined by
Eq.~(\ref{BC}) without resorting to RMT.  In terms of the variable
$z=V\Sigma\lambda_{\rm min}$, the RMT results for the chUE are given
by \cite{Wilk98}
\begin{equation}
  \label{Pmin}
  P(z)=\left\{
    \begin{array}{lll}
      (z/2)\,e^{-z^2/4} & \nu=0, \\
      (z/2)\,e^{-z^2/4}\,I_2(z) & \nu=1, \\
      (z/2)\,e^{-z^2/4}\left[I_2^2(z)-I_1(z)I_3(z)\right] & \nu=2,
    \end{array}\right.
\end{equation}
where $I$ denotes the modified Bessel function.  The RMT results for
the chSE are more complicated and can be found in Ref.~\cite{Berb98}.
The only simple case is $\nu=0$ for which \cite{Forr93}
\begin{equation}
  \label{PminchSE}
  P(z)=\sqrt{\pi/2}\:z^{3/2}\,e^{-z^2/2}\,I_{3/2}(z)\:.
\end{equation}

As we shall show in Sec.~\ref{sec4}, the apparent change from chUE to
chSE symmetry in $P(\lambda_{\rm min})$ is caused by the fact that
$\rho(0)$ (i.e., the extrapolated value) vanishes in a finite volume
for sufficiently weak coupling.  An unambiguous and parameter-free
comparison of numerical data with RMT results is only possible if
$\rho(0)$ and hence $\Sigma$ are nonzero.  In the present work,
however, we are particularly concerned with the region in the
parameter space $(L,\delta)$ where $\Sigma\to0$.  In this case, we are
faced with the problem of how to set the energy scale for
$\lambda_{\rm min}$.  We circumvent this problem by employing three
independent methods of analyzing the data.

(1) We rescale the smallest eigenvalue, both for the numerical data
and for the RMT results, such that $\langle\lambda_{\rm
  min}\rangle=1$.  In this way, the various cases we consider are
characterized only by the shape of the distribution; the ambiguity in
the determination of the scale is avoided.

(2) For each data set, we construct the ratio
\begin{equation}
  \label{ratio}
  r=\frac{\langle\lambda_{\rm min}\rangle}
  {\sqrt{\langle\lambda_{\rm min}^2\rangle
      -\langle\lambda_{\rm min}\rangle^2}}
\end{equation}
which eliminates the energy scale.  The RMT results for this ratio
are
\begin{equation}
  \label{ratioRMT}
  r=\left\{
    \begin{array}{lll}
      1.91306 \quad & {\rm chUE}, & \nu=0, \\
      2.78248 & {\rm chSE}, & \nu=0, \\[2mm]
      2.81978 & {\rm chUE}, & \nu=1, \\
      4.07231 & {\rm chSE}, & \nu=1, \\[2mm]
      3.55611 & {\rm chUE}, & \nu=2, \\
      5.11362 & {\rm chSE}, & \nu=2.
    \end{array}
  \right.
\end{equation}

(3) We fit the data for $P(\lambda_{\rm min})$ to the RMT results of
both the chUE and the chSE (for fixed $\nu$) and apply goodness-of-fit
tests \cite{Eadi71}.  For the determination of the scale, we use the
method of maximum
likelihood.  For the quality of the fit, we use three tests:\\
(3a) A Smirnov-Cram\'er-Von Mises test of the unbinned data (see
Sec.~11.4.1 of Ref.~\cite{Eadi71}).  This test yields a number $NW^2$
which for a good fit is smaller than one ($N$ is the number of
configurations).\\
(3b) A chi-square test of the binned data using equiprobable bins,
with an ``optimal'' number $k$ of bins scaling like $N^{2/5}$ (see
Sec.~11.2.3 of Ref.~\cite{Eadi71}).  For a good fit, the resulting
$\chi^2/(k-1)$ is not much larger than one.\\
(3c) The probability $Q$ that the chi-square should exceed the value
of $\chi^2$ computed in (3b) by chance \cite{Pres86}.  This is given by
$Q=1-P((k-1)/2,\chi^2/2)$ with the incomplete gamma function
$P(a,x)=\int_0^xdt\,t^{a-1}e^{-t}/\Gamma(a)$.  If the probability $Q$
is very small, the theory can be statistically rejected.

Let us emphasize again that in the region where chiral symmetry is
spontaneously broken, $P(\lambda_{\rm min})$ is unambiguously given by
RMT without any free parameter.  Here, however, we explicitly consider
the transition region where $\rho(0)\to0$.  We use the somewhat more
involved methods (1)-(3) to make clear and parameter-independent
statements on the behavior of $P(\lambda_{\rm min})$ in this
transition region.

\section{Results and discussion}
\label{sec4}

We have scanned the parameters $L$ and $\delta$ to reveal the main
properties of the spectrum.  For each parameter set we generated
several thousand random matrices $\widetilde{\mathcal W}$ as described
in Sec.~\ref{sec2}.  For each realization the full spectrum was
computed, and the configurations were grouped according to their
number $\nu$ of zero modes.  The quantities we consider were then
averaged over realizations of $\widetilde{\mathcal W}$.

\begin{figure}
  \begin{center}
    \epsfig{file=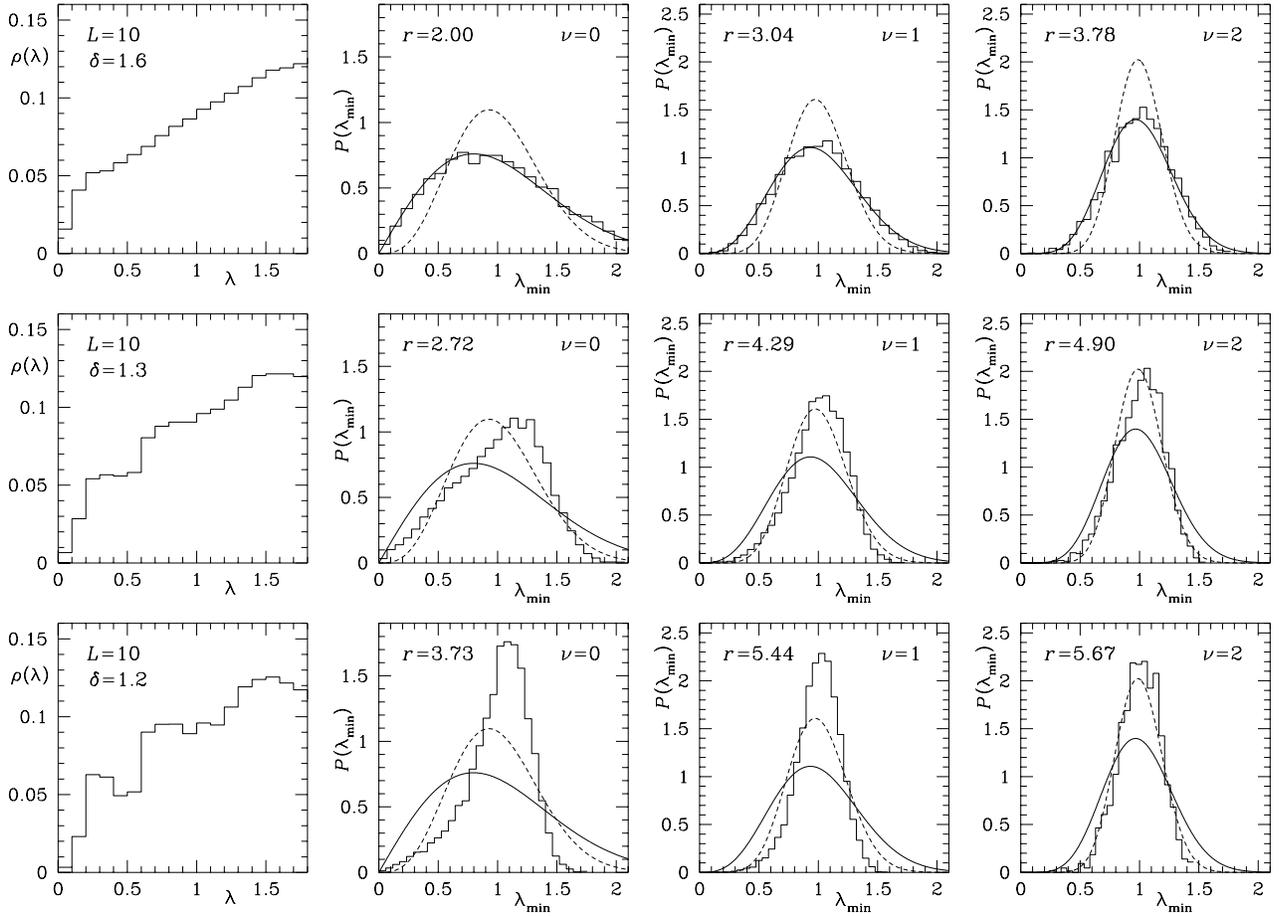,height=122mm} 
  \end{center}
  \vspace*{-2mm}
  \caption{Global spectral density $\rho(\lambda)$ (left column) and
    rescaled distribution of the smallest eigenvalue $P(\lambda_{\rm
      min})$ of $\widetilde{\mathcal W}$ for $L=10$ as a function of
    $\delta$ for $\nu=0,1,2$.  The solid and dotted curves in the
    plots for $P(\lambda_{\rm min})$ are the chUE and chSE
    predictions, respectively.  The quantity $r$ is defined in
    Eq.~(\ref{ratio}).}
  \label{fig1}
\end{figure}

In Fig.~\ref{fig1}, we have fixed the parameter $L=10$ and
varied the parameter $\delta$.  We have plotted the spectral density
$\rho(\lambda)$ in the vicinity of $\lambda=0$ for all $\nu$ combined,
and the rescaled distribution of the smallest positive eigenvalue
$P(\lambda_{\rm min})$ for $\nu=0,1,2$.  (Note that the zero modes are
not included in $\rho(\lambda)$.)  Also given in the figure are the
values of the ratio $r$, see Eq.~(\ref{ratio}).  The results of the
goodness-of-fit tests (3a)-(3b) defined in Sec.~\ref{sec3} are
summarized in Table~\ref{table1}.

The main message of this paper can be read off immediately from these
plots.  The largest value of $\delta=1.6$ corresponds to the strongest
coupling and hence to the largest physical volume.  The extrapolated
value of $\rho(0)$ is clearly nonzero.  Thus, RMT is applicable, and
$P(\lambda_{\rm min})$ agrees very well with the chUE curves for all
three values of $\nu$.  The data also pass the other tests proposed in
Sec.~\ref{sec3}: The ratio $r$ agrees well with that of the chUE, see
Eq.~(\ref{ratioRMT}), and the goodness-of-fit tests, see
Table~\ref{table1}, show that the chUE describes the data well.

On the other hand, the smallest value of $\delta=1.2$ corresponds to
the weakest coupling and hence to the smallest physical volume.  In
this case, the extrapolated value of 

\widetext
\narrowtext

\noindent $\rho(0)$ vanishes, and chiral symmetry is restored.  Thus,
a comparison with RMT becomes meaningless in the microscopic region.
The shape of the distribution $P(\lambda_{\rm min})$ as well as the
ratio $r$ of Eq.~(\ref{ratio}) are now different from the RMT
predictions.  Neither the chUE nor the chSE provide good fits to the
data, as can be seen from the goodness-of-fit tests in
Table~\ref{table1}.  For the intermediate case of $\delta=1.3$, the
value of $\rho(0)$ is neither clearly zero nor clearly nonzero.
Figure~\ref{fig1} and Table~\ref{table1} indicate that already in this
case RMT no longer provides a good description of the data.

\begin{figure}
  \begin{center}
    \epsfig{file=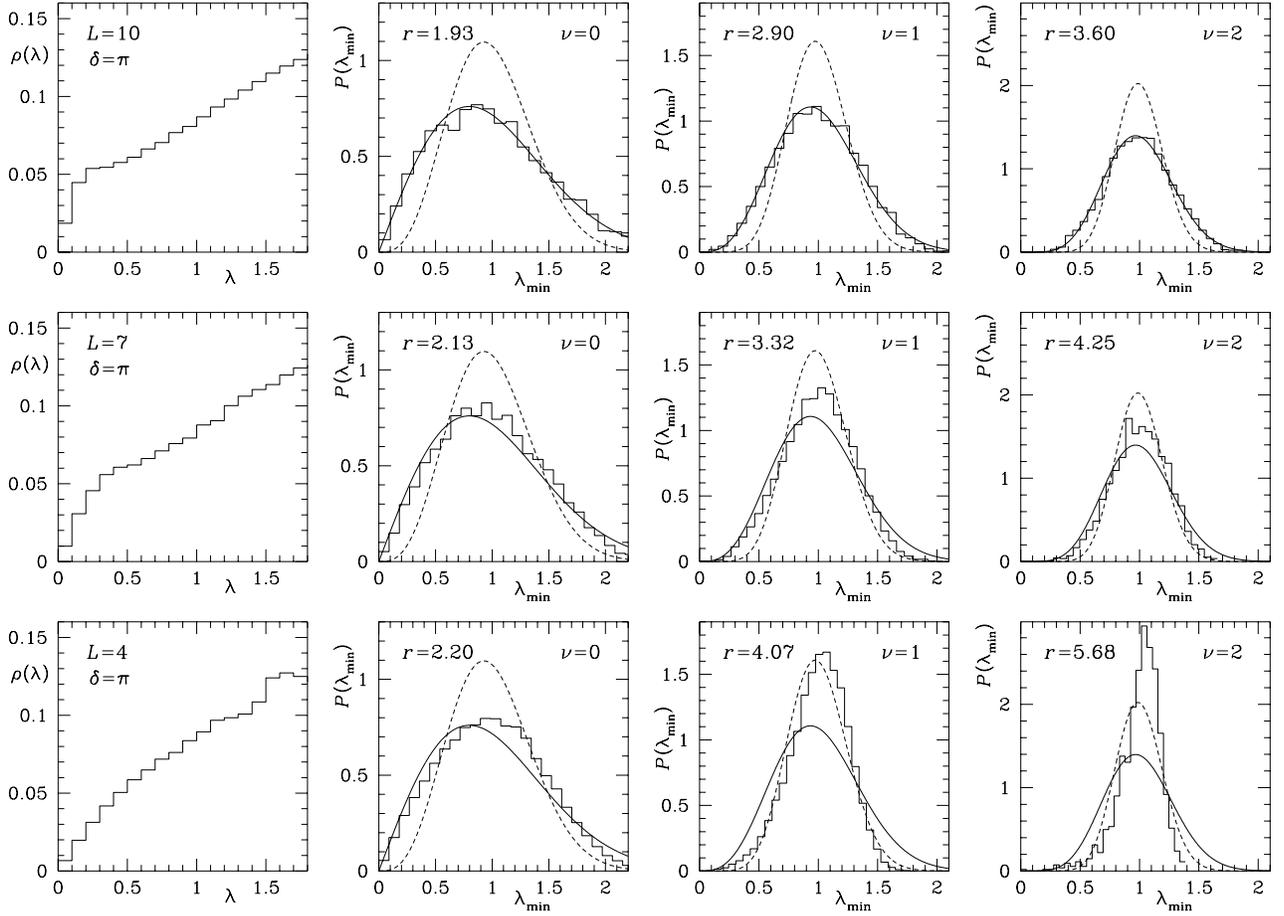,height=122mm} 
  \end{center}
  \vspace*{-2mm}
  \caption{Same as Fig.~\ref{fig1} but keeping $\delta=\pi$ fixed and
    varying $L$.}
  \vspace*{4mm}
  \label{fig2}
\end{figure}

Equivalent conclusions can be drawn from Fig.~\ref{fig2} in which we
exhibit similar plots, but now for a fixed value of $\delta=\pi$
(corresponding to ``strongest coupling'') and $L$ varied from 10 to 4.
The goodness-of-fit tests are also shown in Table~\ref{table1}.
Again, as the physical volume decreases, the extrapolated value of
$\rho(0)$ vanishes, and the agreement between numerical data and RMT
breaks down.  We also observe from Fig.~\ref{fig2} that RMT ceases to
describe the data first for the nonzero values of $\nu$ and then for
$\nu=0$.
%regardless of whether $\delta$ or $L$ is varied to make $\rho(0)$
%vanish.

A remarkable fact, however, is that there are some combinations of the
parameters $L$ and $\delta$ for which the shape of the distribution
$P(\lambda_{\rm min})$ resembles the chSE pre-

\begin{table}
  \begin{center}
    \begin{tabular}{cccllllll}
      $L$ & $\delta$ & $\nu$ & \multicolumn{2}{c}{$NW^2$} &
      \multicolumn{2}{c}{$\chi^2$} & \multicolumn{2}{c}{$Q$} \\
      & & & chUE & chSE & chUE & chSE & chUE & chSE \\\hline\\[-3mm]
      10 & 1.6 & 0 & 0.674 & 25.2 & 1.29 & 14.0 & 0.073 & $10^{-126}$\\
      10 & 1.6 & 1 & 2.49 & 22.25 & 1.47 & 13.9 & 0.008 & $10^{-148}$\\
      10 & 1.6 & 2 & 1.36 & 9.48 & 2.55 & 9.52 & $10^{-8}$ & $10^{-68}$\\
      10 & 1.3 & 0 & 62.8 & 11.7 & 37.6 & 11.8 & 0 & $10^{-126}$\\
      10 & 1.3 & 1 & 73.2 & 8.78 & 45.4 & 7.61 & 0 & $10^{-73}$ \\
      10 & 1.3 & 2 & 11.3 & 1.34 & 9.05 & 3.22 & $10^{-56}$ & $10^{-11}$\\
      10 & 1.2 & 0 & 290 & 103 & 179 & 67.5 & 0 & 0\\
      10 & 1.2 & 1 & 145 & 39.4 & 108 & 29.2 & 0 & 0\\
      10 & 1.2 & 2 & 12.1 & 1.23 & 7.68 & 3.26 & $10^{-37}$ & $10^{-10}$\\
      10 & $\pi$ & 0 & 0.109 & 36.8 & 1.37 & 17.4 & 0.033 & $10^{-176}$\\
      10 & $\pi$ & 1 & 1.20 & 38.9 & 1.88 & 20.3 & $10^{-5}$ & $10^{-254}$\\
      10 & $\pi$ & 2 & 0.268 & 19.4 & 2.10 & 14.4 & $10^{-6}$ & $10^{-131}$\\
      7 & $\pi$ & 0 & 4.88 & 21.9 & 4.12 & 12.6 & $10^{-25}$ & $10^{-127}$\\
      7 & $\pi$ & 1 & 13.8 & 10.7 & 7.17 & 8.98 & $10^{-66}$ & $10^{-90}$\\
      7 & $\pi$ & 2 & 3.71 & 2.40 & 3.82 & 4.22 & $10^{-15}$ & $10^{-18}$\\
      4 & $\pi$ & 0 & 83.6 & 151 & 21.2 & 38.2 & 0 & 0\\
      4 & $\pi$ & 1 & 259 & 25.3 & 116 & 13.3 & 0 & $10^{-276}$\\
      4 & $\pi$ & 2 & 16.6 & 4.97 & 14.0 & 7.14 & $10^{-75}$ & $10^{-31}$ 
    \end{tabular}
    \vspace*{2mm}
    \caption{Results of the goodness-of-fit tests as explained at the
      end of Sec.~\ref{sec3}. Zero means zero to machine precision.} 
    \label{table1}
  \end{center}
\end{table}

\widetext
\narrowtext

\noindent diction.  This effect is particularly pronounced for the
nonzero values of $\nu$, which was also found in Ref.~\cite{FHLW98a}
(for $\nu=1$).  In our data, this is seen in Fig.~\ref{fig1} for
$\delta=1.3$, $\nu=1,2$ and $\delta=1.2$, $\nu=2$, respectively, and
in Fig.~\ref{fig2} for $L=4$, $\nu=1$.  (Recall that the plots of
$P(\lambda_{\rm min})$ are not fits but are obtained by rescaling such
that $\langle\lambda_{\rm min}\rangle=1$.)  However, neither the ratio
$r$ nor the goodness-of-fit tests in Table~\ref{table1} indicate that
the chSE really provides a quantitative description of the data.  The
fact that the chSE seems to work better than the chUE should not be
taken too seriously since neither of the two ensembles is supposed to
be applicable in these cases.  The apparent agreement of the shape of
the curve with the chSE is, for all we know, purely accidental and
should not be interpreted as a transition to symplectic symmetry.

\begin{figure}
  \begin{center}
    \epsfig{file=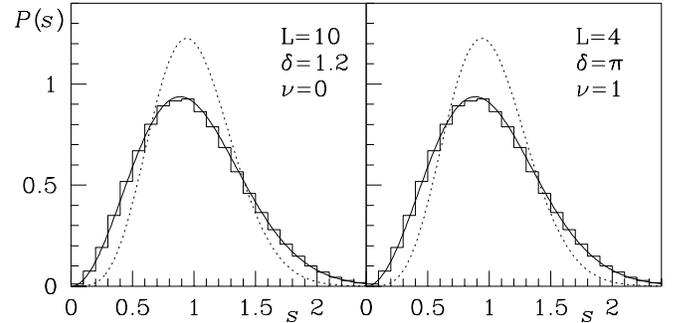,width=\columnwidth}
  \end{center}
%  \vspace*{-5mm}
  \caption{Nearest neighbor spacing distribution $P(s)$ for some of
    the parameter values for which $P(\lambda_{\rm min})$ is not
    described by the chUE.  The histogram represents the data, and the
    solid and dotted curves are the chUE and chSE predictions of
    Eq.~(\ref{Ps}), respectively.}
  \label{fig3}
\end{figure}

We have also constructed $P(s)$ for the same parameter values that
were used in Figs.~\ref{fig1} and \ref{fig2}.  Two examples are shown
in Fig.~\ref{fig3}.  In agreement with the observations of
Ref.~\cite{FHLW98a}, we find that $P(s)$ is always described by the
chUE prediction, even in those cases where $\rho(0)$ vanishes.  This
is consistent with the fact that, even in the chirally symmetric
phase, the bulk spectral correlations of the Dirac operator are still
described by RMT \cite{Pull98}.

\vfill

\section{Conclusions}
\label{sec5}

Let us summarize the picture which has emerged from the study of the
random matrix model.  The Dirac operator of the Schwinger model has
the symmetries of the chUE, and our model has the same symmetries, the
only essential difference being the replacement of the U(1) gauge
fields by random phases.  The bulk spectral correlations of the Dirac
operator, measured by $P(s)$, are described by the chUE result for all
values of the parameters, regardless of whether or not $\rho(0)$ is
nonzero. This is consistent with the findings of Ref.~\cite{FHLW98a}.
The microscopic spectral correlations, here measured by
$P(\lambda_{\rm min})$, are given by the chUE predictions only for
those values of the parameters for which $\rho(0)>0$.  By changing the
parameters one can decrease the physical volume and cause $\rho(0)$ to
vanish.  The small eigenvalues are then no longer described by RMT.
In this case, however, there exist some special values of the
parameters for which the shape of the distribution of the smallest
eigenvalue resembles that of the chSE.  However, other measures such
as the ratio $r$ of Eq.~(\ref{ratio}) and the various goodness-of-fit
tests clearly show that the chSE does not describe the data.  Thus,
the apparent agreement with the symplectic symmetry case is an
artifact of chiral symmetry restoration in a finite volume and should
be regarded as accidental.

There is a simple lesson to be learned from our analysis.  If one
wants to use random matrix methods to analyze data from lattice
simulations, one has to make sure that one is working in a regime in
which RMT is applicable.  For the bulk spectral correlations, this is
never really an issue.  However, for RMT to describe the small
Dirac eigenvalues it is necessary that $\rho(0)>0$, i.e., that chiral
symmetry is spontaneously broken.

\acknowledgments

This work was supported by the Deutsche Forschungsgemeinschaft (DFG),
the US Department of Energy (DOE), and the RIKEN BNL Research Center
(RBRC).  We would like to thank A. Sch\"afer for many stimulating
discussions throughout the course of this study.  We also thank the
authors of Ref.~\cite{FHLW98a} for helpful communication and M.
G\"ockeler, J.J.M.  Verbaarschot, and H.A.  Weidenm\"uller for
valuable comments.

\widetext
\end{document}